\documentstyle[prl,aps]{revtex}
\begin{document}
\draft
\tighten
\twocolumn[\hsize\textwidth\columnwidth\hsize\csname @twocolumnfalse\endcsname
\title{Strongly Localized Electrons in a Magnetic Field: Exact Results 
on Quantum Interference and Magnetoconductance}
\author{Yeong-Lieh Lin$^{1,2}$ and Franco Nori$^{1}$}
\address
{$^{1}$ Department of Physics, The University of Michigan, 
Ann Arbor, Michigan 48109-1120 \\
$^{2}$Department of Physics, West Virginia University, 
Morgantown, West Virginia 26506-6315\,\cite{lyl} }
\date{\today}
\maketitle
\begin{abstract}
We study quantum interference effects 
on the transition strength
for strongly localized electrons 
hopping on 2D square and 3D cubic 
lattices in a magnetic field ${\bf B}$. 
In 2D, we
obtain {\em closed-form} expressions for the tunneling probability 
between two {\em arbitrary} sites by exactly summing the
corresponding phase factors of {\em all} directed paths connecting
them. An {\em analytic} expression for the magnetoconductance, as an 
{\em explicit} function of the magnetic flux,  is derived. 
In the experimentally important 3D case, we show how the interference 
patterns and the small-${\bf B}$ 
behavior of the magnetoconductance  
vary according to the orientation of ${\bf B}$.
\end{abstract}
\vskip2pc]
\vspace{-1in}
\narrowtext

Quantum interference (QI) effects 
between different electron paths in disordered electron systems 
have been a subject of intense 
study\cite{weak,1,2,3,4,5,6,7}
because they play an important role in quantum transport; for instance, 
the QI of  closed paths is central to {\em weak}-localization
phenomena\cite{weak}.  Recently, a growing interest exists on the effects of 
a magnetic field on {\em strongly localized} electrons with 
variable-range hopping (VRH) where striking QI phenomena have 
been observed in mesoscopic and macroscopic 
insulating materials. 
This strongly localized regime\cite{1,2,3,4,5,6,7} 
is less well-understood than 
the weak-localization case. 
Deep in the insulating  regime, the major mechanism for transport 
is thermally activated hopping between the localized sites. 
In the VRH regime, localized electrons hop a long distance (the lower the 
temperature is, the further away the electron 
tunnels) in order to  find a localized site of close energy. 
The conductance of the sample is governed by one 
critical phonon-assisted hopping event\cite{1}. 
During this critical tunneling process, 
the electron traverses many other 
impurities since the hopping length is typically many times 
larger than the localization length. It is important to emphasize that 
the electron preserves its phase memory 
while encountering these intermediate scatterers. 
This elastic multiple-scattering 
is the origin of the QI effects associated with a single hopping event 
between the initial and final sites. The tunneling 
probability of one distant hop is therefore determined by the interference of 
many electron paths between the initial and final sites\cite{1,2,3,4,5,6}.

In this paper we investigate the QI of strongly localized electrons
by doing {\em exact} summations over {\em all} 
directed paths between two {\em arbitrary} sites. 
For electrons propagating on a square lattice under a uniform potential, 
we derive an exact {\em closed-form} expression for the sum-over-paths. 
We also obtain an explicit formula for an experimentally important case, 
much less studied theoretically so far: the interference between paths on 
a 3D cubic lattice. In the presence of impurities, by computing the 
moments of the tunneling probability and employing the replica method, 
we derive a compact {\em analytic} result for the 
magnetoconductance (MC), which is applicable in any dimension. 
Our explicit field-dependent expression for the MC provides 
a precise description of the MC, including the low and high 
field limits. The period of oscillation of the MC is found to 
be equal to $hc/2e$.  Also, a {\em positive} MC is clearly 
observed when turning on the field ${\bf B}$. 
When the strength of ${\bf B}$ reaches a certain value, $B_{{\rm sa}}$, 
which is inversely proportional to twice the hopping length, 
the value of the MC becomes saturated. At very small fields, for two sites 
diagonally separated a distance $r$, the MC behaves as: $rB$ for 
quasi-1D systems, $r^{3/2}B$ in 2D with ${\bf B}=(0,0,B)$, and 
$rB$ [$r^{3/2}B$] in 3D with ${\bf B}$ parallel [perpendicular] 
to the $(1,1,1)$ direction. The general expressions presented here
(i) {\em contain, as particular cases}, several QI 
results\cite{1,2,3,4,5,6,7}  
derived during the past decade 
(often by using either numerical or approximate methods),
(ii) include QI to arbitrary points $(m,n)$, 
instead of only diagonal sites $(m,m)$, 
(iii) focus on 2D {\em and} 3D lattices, 
and (iv) can be extended to also include {\em backward} excursions 
(e.g., side windings) in the directed paths.

Exact results in this class of directed-paths problems are valuable; and, for 
instance, can be useful to study other systems: (1) directed polymers 
in a disordered substrate, (2) interfaces in 
2D, (3) light propagation in random media, 
and (4) charged bosons in 1D.

To study the magnetic field effects on the tunneling probability 
of strongly localized electrons, we start from the tight-binding 
Hamiltonian
$H=W \sum_{i} c_{i}^{\dag}c_{i} + V \sum_{\langle ij \rangle}
c_{i}^{\dag}c_{j}e^{i A_{ij}},$
where $V/W \ll 1$ and $A_{ij}=2\pi \int_{i}^{j}{\bf A}{\cdot}d{\bf l}$ is 
$2\pi$ times the line 
integral of the vector potential along the bond from $i$ to $j$ in 
units of $\Phi_{0}=hc/e$. 
Consider two states, localized at sites $i$ and $f$ which 
are $r$ bonds apart, and the shortest-length paths 
(with no backward excursions) connecting them; i.e., the 
{\em directed-path model}.  By using a locator 
expansion, the Green's function (transition amplitude) 
between these two states can be expressed as
$T_{if}=W(V/W)^{r}S^{(r)},$ with 
$S^{(r)}= \sum_{\Gamma}\exp(i\Phi_\Gamma),$ 
where $\Gamma$ runs over all directed paths of $r$ steps 
and $\Phi_{\Gamma}$ is the sum over phases of the bonds on the path. 
This directed-path model provides an excellent approximation to $T_{if}$ 
in the extremely localized regime\cite{1,2,3,4,5,6,7} 
since higher-order contributions involve terms propotional to 
$W(V/W)^{r+2l}$ ($l \geq 1$), which are negligible because $(V/W)^2$ 
is very small. 
Quantum interference, contained in 
$S^{(r)}$, arises because the phase factors of different paths 
connecting the initial and final sites interfere with each other. 
In this work, we focus on: $(i)$ the computation of 
the essential QI quantity $S^{(r)}$, $(ii)$ the derivation of the 
MC in the disordered case, and $(iii)$ the study of the full behavior of the 
MC---including the scaling in the low-field limit and 
the occurrence of saturation. 
It is important to keep in mind that the effect of a 
magnetic field on the MC follows the behavior of $S^{(r)}$.

{\em Quantum Interference on a 2D 
square lattice.---}Let us choose $(0,0)$ to
be the initial site and focus on sites $(m,n)$ with 
$m,n \geq 0$. For forward-scattering
paths of $r$ steps, which exclude backward excursions, 
ending sites $(m,n)$
satisfy $m+n=r$. Let $S_{m,n}$ (=$S^{(r)}$) be the sum over all directed
paths of $\,r\,$ steps on which an electron can hop from the origin to 
$(m,n)$, each one weighted by its corresponding phase factor.
Employing the symmetric gauge ${\bf A}=(-y,x)B/2$, 
and denoting the flux through an elementary plaquette 
(with an area equal to the square of the average distance, which is 
typically equal to or larger than the localization length, 
between two impurities) by $\phi/2\pi$, 
it is straightforward to construct the recursion relation: 
$S_{m,n}=e^{-in\phi/2}\,S_{m-1,n}\,+\,e^{im\phi/2}\,S_{m,n-1}.$ 
The factors in front of the $S$'s account for the presence of 
the magnetic field. 
Enumerating the recursion relations for $S_{k_{n},n}$ 
($k_{n}=m-1,\ldots,0$) successively and using $S_{0,n}=1$, 
we obtain the following relation 
$S_{m,n}=\sum_{k_n=0}^{m}\,e^{ik_n\phi/2}\,
e^{-i(m-k_n)n\phi/2}\,S_{k_n,n-1}.$ This equation states that 
the site $(m,n)$ can be reached by moving one step upward 
from sites $(k_{n},n-1)$ with $0 \leq k_{n}\leq m$, acquiring the 
phase $ik_{n}\phi/2$; then traversing $m-k_n$ steps from $(k_n,n)$ 
to $(m,n)$, each step with a phase $-in\phi/2$. 
By applying the above relation 
recursively and utilizing $S_{m,0}=1$, 
$S_{m,n}$ for $m,n \geq 1$ can be written as 
$S_{m,n}(\phi)=\exp(-i mn \phi/2)\,L_{m,n}(\phi)$, and
\vspace{-0.09in} 
\begin{equation}
L_{m,n}(\phi)=\sum_{k_n=0}^{m}
\sum_{k_{n-1}=0}^{k_{n}}\cdots\sum_{k_1=0}^{k_2}
\,e^{i(k_1+\cdots+k_{n-1}+k_{n})\phi}. 
\end{equation}

\vspace{-0.09in}
\noindent
Notice that each term in the summand corresponds to the overall phase 
factor associated with a directed path. When $\phi =0$,
$\ S_{m,n}(0)=C_{m}^{r}\equiv N$ 
is just the total number of $r$-step paths between $(0,0)$ 
and $(m,n)$.

After some calculations we obtain one of our main results, 
a very compact and elegant closed-form expression for $S_{m,n}(\phi)$:
\vspace{-0.09in}
\begin{equation}
S_{m,n}(\phi)=\frac{F_{m+n}(\phi)}{F_{m}(\phi)F_{n}(\phi)},
\ \ \ F_{m}(\phi)=\prod_{k=1}^{m}\sin\frac{k}{2}\phi.
\end{equation}

\vspace{-0.09in}
\noindent
Notice that the symmetry $S_{m,n}=S_{n,m}$ [apparent in Eq.~(2)] is due
to the square lattice geometry. In the very-low-flux limit $\phi \ll 1$, 
the logarithm of
$S_{m,n}$, calculated {\em exactly} to order $\phi^2$ 
(and omitting $\ln N$), is
\vspace{-0.09in}
\begin{equation}
\ln S_{m,n}(\phi)=-\frac{1}{24}m\,n\,(m+n+1)\,\phi^{2},
\end{equation}
and thus we obtain the familiar\cite{1} harmonic 
shrinkage of the wave function.

$S_{m,m}$ has the richest interference effects because the
number of paths ending at $(m,m)$ and the areas they enclose are both 
the largest. We therefore examine
more closely the behavior of the quantities 
$I_{2m}(\phi)\ \equiv\ S_{m,m}(\phi)\ =\ \prod_{k=1}^{m}\,
(\sin\frac{m+k}{2}\,\phi)/(\sin\frac{k}{2}\,\phi).$ 
$I_{2m}(\phi)$ obeys the following properties:
(i) $2\pi$ ($4\pi$) periodicity in $\phi$ for even (odd) $m$, 
(ii) $I_{2m}(2\pi-\phi)=I_{2m}(\phi)$ for $0 \leq \phi \leq \pi$ 
with $m$ even, and 
(iii) $I_{2m}(2\pi\pm\phi)=-I_{2m}(\phi)$ for 
$0 \leq \phi \leq 2\pi$ with $m$ odd. Furthermore, 
the zeros of $I_{2m}(\phi)$ are given
by $\phi=2\pi\,s/t$, for $\frac{m+1}{n+1} \leq t \leq \frac{2m}{2n+1}$,
with $0 \leq n \leq \frac{m-1}{2} $, and the $s$'s are prime to each allowed
$t$. From the physical viewpoint, these flux values produce the complete 
cancelation of all phase factors 
(i.e., {\em fully destructive interference}). Indeed, 
as the magnetic field is turned on, $I_{2m}(\phi)$ rapidly drops to 
its first zero at $\phi/2\pi=1/2m$ and then shows 
many small-magnitude fluctuations around zero.

{\em Effects of disorder.---}To incorporate the effects of random 
impurities, we now replace the on-site energy part (first term in $H$)
by $\sum_{i}\epsilon_{i}c_{i}^{\dag}c_{i}$. Now the $\epsilon_{i}$ are 
independent random variables which can take two values: $+W$ with
probability $\mu$ and $-W$ with probability $\nu$, where $\mu+\nu=1$.
Due to disorder, the transition amplitude becomes 
$T_{if}=W(V/W)^rJ_{m,n},$ with 
$J_{m,n}=\sum_{\Gamma}[\,\prod_{j\in \Gamma}
(-W/\epsilon_j)\,]e^{i\Phi_{\Gamma}}.$ 
For all directed paths ending at $(m,n)$, electrons traverse $r$
sites (the initial site is excluded). Each site visited now
contributes an additional multiplicative factor of either $+1$ or $-1$
to the phase factor. Therefore, for a given path $\Gamma$, the
probability for obtaining 
$\pm e^{i\Phi_{\Gamma}}$ is $[(\mu+\nu)^{r}\pm(\mu-\nu)^{r}]/2$. 
By exploiting Eq.~(1), we derive a  general 
expression for the disorder average of the tunneling probability 
(i.e., the transmission rate) as  
$\langle\,|J_{m,n}(\phi)|^{2}\,\rangle =N
+(\mu-\nu)^{2r}[S^{2}_{m,n}(\phi)-N],$ where $\langle\cdots\rangle$ denotes 
averaging over all possible configurations of impurities.
It is important to stress that the {\em conductivity} between 
$i$ and $f$ is proportional to 
$\langle\,|J_{m,n}(\phi)|^{2}\,\rangle$\cite{1,2,3,4}.

For the most studied case so far: $\mu=\nu=1/2$, we can obtain 
analytical expressions for the moments 
$ \langle \, |J_{m,n}(\phi)|^{2p}\rangle $ 
for any value of $p$. 
In general, $\langle\, |J_{m,n}(\phi)|^{2p} \rangle$ consist of terms 
involving $N^{k}$ with $k=1,\ldots, p.$ Hereafter, we omit the subscripts in 
$J$ and $S$, and focus on the leading terms 
($\propto N^{p}$), since they provide the most significant contribution 
to the moments, and hence the MC, when $N$ is large. 
Recall that $S(0)=N$, therefore 
we need to consider all terms involving 
$S^{2k}(2\phi)\,N^{p-2k}$ in $\langle\, |J(\phi)|^{2p} \,\rangle$. 
We derive $\langle\, |J(0)|^{2p} \,\rangle =(2p-1)!!\,N^{p}$, and
\begin{equation}
\langle\, |J(\phi)|^{2p} \,\rangle = p!\,N^{p}\,\left\{\sum_{k=0}^{\infty}
\frac{(2k)!\,C^{p}_{2k}}{(2^k\,k!)^{2}}\,
\left[\frac{S(2\phi)}{N}\right]^{2k}\right\}.
\end{equation}

Using these equations and employing the replica method, we obtain the 
log-averaged MC $L_ {MC}\equiv \langle\, \ln|J(\phi)|^{2} \,\rangle 
-\langle\, \ln|J(0)|^{2} \,\rangle$. 
The typical MC of a sample, 
$G(\phi)=\exp(\langle \ln |J(\phi)|^2\rangle)$, is then given by
\begin{equation}
\frac{G(\phi)}{G(0)}=\exp(L_{MC})=1
+\sqrt{1-\left[\frac{S(2\phi)}{N}\right]^2}\,.
\end{equation}
Eq.~(5) is one of our main results.
It provides a compact closed-form expression 
for the MC, as an {\em explicit} function of the magnetic flux. 
From Eq.~(5) it becomes evident that 
a magnetic field leads to an increase in the {\em positive} MC: 
$G(\phi)/G(0)$ increases from $1$ to a saturated value $2$ 
(since $S(2\phi)$ decreases from $N$ to $0$) when the flux 
is turned on and increased. $G(\phi)=2G(0)$ at the field $\phi$ that satisfies 
$S(2\phi)=0$. Furthermore, 
it is clear that the MC varies {\em periodically} with the 
magnetic field and the 
periodicity in the flux is equal to $hc/2e$.

It is illuminating to draw attention to the 
close relationship between the behaviors of 
$I_{2m}(2\phi)=S_{m,m}(2\phi)$ and the corresponding $G(\phi)$. When 
$\phi=0$, $(I_{2m}(0)/N)^2=1$, which is the {\em largest} value 
of $(I_{2m}(2\phi)/N)^2$ as a function of $\phi$, and the MC is 
equal to the {\em smallest} value $G(0)$. When the magnetic field is 
increased from zero, $(I_{2m}(2\phi)/N)^2$ quickly approaches 
(more rapidly as $m$ becomes larger) its {\em smallest} 
value, which is zero, at $\phi/2\pi=1/4m$. 
At the same time, the MC rapidly increases to the {\em largest} 
value $2G(0)$. The physical implication of this is clear: fully constructive 
(destructive) interference in the case without disorder leads to  the 
smallest (largest) hopping conduction in the presence of disorder.  
Moreover, when $m$ (the system size is $m\times m$) is large, 
$G(\phi)/G(0)$ remains in the close vicinity of $2$ for 
$\phi/2\pi>1/4m$ in spite of the strong very-small-magnitude 
fluctuations of $I_{2m}(2\phi)/N$ around zero.
 
The saturated value of the 
magnetic field $B_{{\rm sa}}$ [i.e., the first field that 
makes $G(\phi)=2G(0)$] is 
inversely proportional to twice the hopping length: the larger the system is, 
the smaller $B_{{\rm sa}}$ will be. In other words, as soon as the 
system, with hopping distance $r=2m$, is penetrated by a total flux 
of $(1/2r)\times(r/2)^2=r/8$ (in units of $\Phi_{0}$), 
the MC reaches the saturation value $2G(0)$. 

To examine the behavior of the MC in the low-flux limit, we first define 
the relative MC, $\Delta G(\phi)\equiv[G(\phi)-G(0)]/G(0).$
Since $\ln I_{r}=-r^{2}(r+1)\phi^2/96$ and 
$\ln S_{r-1,1}=-(r^{2}-1)\phi^2/24$ from Eq.~(3), 
it follows then that, for very small fields, in 2D 
$\Delta G(\phi) \simeq \sqrt{3}r^{3/2}\phi/6$ 
and in ladder-type quasi-1D structures 
$\Delta G(\phi) \simeq \sqrt{3}\,r\,\phi/3.$

Our results for the MC are 
in good agreement with experimental 
measurements.  For instance, a positive MC is observed in 
the VRH regime of both macroscopic In$_2$O$_{3-x}$ 
samples and compensated n-type CdSe\cite{exp1}. 
Moreover, saturation in 
the MC as the field is increased is also reported in Ref.~\cite{exp1}.

The result for $G(\phi)$ presented in this work is consistent with 
theoretical studies based on an independent-directed-path 
formalism 
and a random matrix theory of the transition strengths\cite{7}. 
The advantages of our results lie in that they provide $(i)$ an 
explicit expression for the MC as a function of the magnetic field, 
and thus a straightforward 
determination of the period of oscillation, 
$(ii)$ explicit scaling behaviors 
(i.e., the dependence on the hopping 
length as well as  the orientation and strength of the field) 
of the low-flux MC in quasi-1D, 2D and 
3D systems, and $(iii)$ allow a quantitative comparison with 
experimental data. 
It is important to emphasize that our analytic result for the MC is 
equally applicable to {\em any dimension}, since the essential 
ingredient in our expressions is the QI quantity $S^{(r)}$, which 
takes into account the dimensionality.

{\em Quantum Interference  and the small-field magnetoconductance 
on a 3D cubic lattice.---}Let ${\cal S}_{m,n,l}$ 
($=S^{(r)}$ in 3D) be the sum over all phase factors associated 
with directed paths of $m+n+l\,(=r)$ steps 
along which an electron may hop from (0,0,0) to the site $(m,n,l)$ 
with $m$, $n$ and $l \geq 0$. 
The vector potential of $(B_{x},B_{y},B_{z})$ can be written as 
${\bf A}=(zB_y-yB_z, \, xB_z-zB_x, \, yB_x-xB_y )/2$.  Also, 
$a/2\pi$, $b/2\pi$ and $c/2\pi$ represent the three fluxes through the 
respective elementary plaquettes on the $yz$-, $zx$- and $xy$-planes. 
To compute ${\cal S}_{m,n,l}$, we start from the following 
recursion relation ${\cal S}_{m,n,l}=\sum_{p=0}^{m} 
\sum_{q=0}^{n}A_{p,q,l\rightarrow m,n,l}\,
e^{i(qa-pb)/2}\,{\cal S}_{p,q,l-1},$
where $A_{p,q,l\rightarrow m,n,l}$ is the sum over all directed 
paths starting from $(p,q,l)$ and ending at $(m,n,l)$. 
The physical meaning of this relation is clear: the site $(m,n,l)$ is 
reached by taking one step from $(p,q,l-1)$ to $(p,q,l)$, acquiring 
the phase $i(qa-pb)/2$, then traversing from $(p,q,l)$ to $(m,n,l)$ 
on the $z=l$ plane. We find that
$A_{p,q,l\rightarrow m,n,l}=
\exp\{ i [(m-p)(lb-qc)+(n-q)(pc-la)]/2 \}S_{m-p,n-q}(c).$ By 
applying the above equation $l$ times, we obtain a general formula 
of ${\cal S}_{m,n,l}$ for $m,n,l \geq 1$ 
in terms of the fluxes $a$, $b$ and $c$ as
\vspace{-0.09in}
\begin{equation}
{\cal S}_{m,n,l}(a,b,c)=e^{-i(nla+lmb+mnc)/2}\,{\cal L}_{m,n,l}(a,b,c),
\end{equation}
\vspace{-0.3in}
\begin{eqnarray}
{\cal L}_{m,n,l}\!&=&\!\left\{\!\prod_{j=1}^{l}\!
\left[ \sum_{p_j=0}^{p_{j+1}}\sum_{q_j=0}^{q_{j+1}}\!e^{i[q_ja+(m-p_j)
b+p_j(q_{j+1}-q_j)c]}\right. \right. \nonumber  \\
& & \mbox{}\times \left. \left. 
L_{p_{j+1}-p_j,q_{j+1}-q_j}(c)\right]\right\}\,L_{p_1,q_1}(c), 
\end{eqnarray}

\vspace{-0.09in}
\noindent
with $p_{l+1}\equiv m$, $q_{l+1}\equiv n$, and the $L_{p,q}(c)$'s are 
defined as in Eq.~(1). In the absence of the flux, 
${\cal S}_{m,n,l}=(m+n+l)!\,/\,m!\,n!\,l! \equiv {\cal N}$ 
gives the total number of $r$-step paths 
connecting $(0,0,0)$ and $(m,n,l)$. 
In the very-low-flux limit, {\em exactly} calculated to second order 
in the flux and omitting the term $\ln{\cal N}$, the 3D analog of the 
harmonic shrinkage of the wave function becomes
\vspace{-0.09in}
\begin{eqnarray}
\ln{\cal S}_{m,n,l}&=&-\frac{1}{24}\left[nla^2+lmb^2+mnc^2 \right. \\
& &\mbox{}\left. +m(lb-nc)^2\!+\!n(mc-la)^2\!+\!l(na-mb)^2\right]. \nonumber
\end{eqnarray}

\vspace{-0.09in}
\noindent
In order to see how the interference patterns and the MC vary according to 
the orientation of the applied field, 
we now examine two special cases: 
${\bf B}_{\parallel}=(1,1,1)(\phi/2\pi)$ and 
${\bf B}_{\perp}=(1/2,1/2,-1)(\phi/2\pi)$, namely, fields parallel and 
perpendicular to the $(1,1,1)$ direction.  We find that 
their $S_{m,m,m}$, designated respectively by ${\cal I}_{r}^{\parallel}$ 
and ${\cal I}_{r}^{\perp}$ (where $r=3m$), exhibit quite different behaviors. 
Furthermore, they are insensitive to 
the commensurability of $\phi$, unlike the case on a square lattice. 
Physically, this can be understood because paths have a higher probability 
of crossing (and thus interfering) in 2D than in 3D;
thus making QI effects less pronounced in 3D than in 2D. 
A similar situation occurs classically (e.g., multiply-scattered light 
in a random medium).   For very small $\phi$, 
$\ln {\cal I}_{r}^{\parallel} =-r^{2}\phi^{2}/72$ and 
$\ln {\cal I}_{r}^{\perp} =-r^{2}(r+1)\phi^{2}/144$. 
The 3D behavior of $\Delta G(\phi)$ thus become clear: $\simeq r\phi/3$ for 
${\bf B}_{\parallel}$ and $\simeq \sqrt{2}r^{3/2}\phi/6$ 
for ${\bf B}_{\perp}$. These results can be 
interpreted as follows: the effective area exposed to ${\bf B}_{\parallel}$ 
is smaller ($\sim r$), similar to our quasi-1D case with 
$\Delta G(\phi)\propto r\phi$; while the effective area exposed to 
${\bf B}_{\perp}$ is larger ($\sim r^{3/2}$), thus closer to the 2D 
case with $\Delta G(\phi)\propto r^{3/2}\phi$.

{\em Average of the magnetoconductance over angles.---\ }
In a macroscopic 
sample, the conductance may be determined by {\em a few} 
(i.e., more than one) critical hopping events. 
As a result of this, the observed MC of the 
whole sample should then be the average of the MC associated with 
these critical hops\cite{AM}. Thus, in 3D systems, 
it is also important to take into account 
the randomness of the angles between the hopping direction and the 
orientation of the applied magnetic field. We therefore  
consider the following 
picture: the ending site of all hopping 
events (with the same hopping length $r$) is located at the 
body diagonal $(r/3,r/3,r/3)$ and the magnetic field can be adjusted 
between the parallel and perpendicular directions with respect to 
the vector ${\bf d}=(1,1,1)$. Our interest here is in 
the MC averaged over angles, 
denoted by $\overline{\Delta G}$, in the low-field limit. 
From Eq.~(8), we obtain 
$\Delta G(B)=(2\pi/3\sqrt{3})rB(1+r\sin^2\omega)^{1/2}$, 
where $B$ is the magnitude of the field and 
$\omega$ is the angle between ${\bf B}$ and ${\bf d}$. 
By averaging over the angle $\omega$, we obtain
\vspace{-0.09in}
\begin{equation}
\overline{\Delta G}(B)=\frac{4}{3\sqrt{3}}\,r\sqrt{r+1}\,B\,
E\!\left(\frac{\pi}{2},\frac{\sqrt{r}}{\sqrt{r+1}}\right)\,,
\end{equation}

\vspace{-0.09in}
\noindent
where $E(\pi/2,\sqrt{r}/\sqrt{r+1})$ is the complete elliptic 
integral of the second kind. 
When $r$ is large, $E\simeq 1$ and 
we therefore have 
$\overline{\Delta G}(B)\simeq(4/3\sqrt{3})\,r^{3/2}\,B.$ 
This  means that the dominant contribution to the MC stems from the 
critical hop which is perpendicular to the field. This is understandable 
through our earlier observation that the effective area enclosed by the 
electron is largest when ${\bf B}$ is perpendicular to ${\bf d}$. 
From the above analysis, we conclude that in 3D macroscopic samples 
the low-field MC should in principle behave as $r^{3/2}B$.

Finally we briefly address five issues.  First, in addition to the 
dominant terms, we have also 
obtained the second-order contribution to the moments 
($\propto N^{p-1}$) and the MC ($\propto 1/N$). The principal features in 
the behavior of the MC are not significantly modified: only the magnitude 
of the positive MC, including the saturation value, is slightly increased 
(though negligibly small).
Second, we have also studied another model of disorder: 
$\epsilon_{i}$ uniformly distributed between $-W/2$ and $W/2$. The result 
for the MC remains the same.
Third, returns to the origin become important 
for less strongly localized electrons, and their QI effects\cite{9} 
can be incorporated in our approach. 
Fourth, the main limitations of our study are the following: no inclusion of
spin-orbit scattering effects (for this see, e.g., Refs.~[7,8] 
and references therein), and no explicit inclusion of the correlations 
between crossing paths, as discussed in Refs.~[4,7]. 
However, these correlations 
are negligible when spin-orbit scattering is 
present\cite{6}. Fifth, our result for $\Delta G(B)$ in 3D can be 
tested experimentally by measuring the MC of bulk samples for varying 
orientations of the field.

In summary, we present an investigation of QI phenomena and the 
magnetic-field effects on the MC for 2D and 3D systems in the VRH regime. 
We provide exact and explicit closed-form results 
for the forward-scattering paths which, in the strongly localized regime, 
give the dominant contribution to the hopping conduction.

We are very grateful to B.L. Altshuler, M. Kardar, E. Medina, Y. Meir, 
Y. Shapir, and J. Stembridge 
for their useful suggestions.

\end{document}